# On Lepton Flavor Mass-Degeneracy-Deviation Hierarchies and QD-Neutrino Mass

## E. M. Lipmanov

40 Wallingford Road # 272, Brighton MA 02135, USA

**Abstract**

Two guiding new phenomenological flavor and flavor-electroweak ideas are expounded in this paper: (I) Oppositeness relation between neutrino and Charged Lepton (CL) Mass-Degeneracy-Deviation (MDD) quantities. With inputs from the CL mass and neutrino oscillation data, it enables two independent estimations of the Quasi-Degenerate (QD) neutrino masses $m_\nu \cong$ (0.11-0.30) eV and $m_\nu \cong$ (0.11-0.16) eV. Small value of the neutrino oscillation hierarchy parameter is a consistency condition for the two opposite solutions. (II) Essential connection between lepton mass hierarchies and low energy electroweak coupling constants. It enables an independent estimation of the QD-neutrino mass $m_\nu \cong (0.11-0.14)$ eV, and renders three sequential CL copies into one flavor system with the elementary charge encoded in the three-flavor MDD-hierarchy. An exact relation between the fine structure constant at $Q^2=0$ and the experimental CL mass-ratio parameter $\alpha_o \equiv e^{-5}$ is observed. Experimental evidence is suggesting that $\alpha_o$ has physical meaning of a dimensionless universal parameter that links flavor and electroweak quantities in accord with the guiding idea (II).

## 1. Introduction

Unlike the one-generation Standard Model, flavor mass physics is still an empirical frontier-sector of physics. With that, the empirical forms of the flavor mixing matrixes are much more elaborated in the low energy phenomenology (Cabibbo-Kobayashi-Maskawa quark mixing matrix and Pontecorvo-Maki-Nakagawa-Sakata neutrino mixing one) than the particle mass-ratios as the other parts of the low energy particle mass matrixes. While the particle mixing matrixes are important for actual calculations, the mass ratios may be more indicative of new flavor physics. There are still several basic problems in low energy lepton flavor mass and neutrino phenomenology, e.g. (a) What is the pattern of the neutrino masses in comparison with the mass patterns of the CL and quarks? (b) What is the neutrino mass scale? (c) Are the main low energy characteristics of the lepton mass spectra in flavor physics - the hierarchies of masses $m_n$ and mass-ratios $x_n \equiv (m_{n+1}/m_n)$ - connected to known low energy quantities of basic **p**hysics and what may be the connection? New phenomenological ideas in favor of a definite type of neutrino mass pattern and neutrino mass scale may be stimulating. In that spirit, I suggested in ref.[1] an opposite relation between the neutrino and CL MDD $(x_n-1)$-quantities; it singles out the widely discussed in the literature possibility of QD-neutrino mass type, determines the absolute QD-neutrino masses in terms of the oscillation mass-squared differences, and leads to a small [2] solar-atmospheric hierarchy parameter $\Delta m^2_{sol}/\Delta m^2_{atm}$ of the QD-neutrinos, probably dynamically related to electroweak physics.

Among the known elementary particles, the CL and quarks are Dirac particles with divergent hierarchical flavor mass patterns, while the neutrinos are the only particles that may be of Majorana nature and quasi-degenerate. Against this data **b**ackground, two guiding compatible new flavor physics ideas are **e**xpounded in this paper: (I) Oppositeness relation between neutrino and CL MDD-quantities; it suggests QD-neutrinos with a small solar-atmospheric hierarchy parameter as a consistency condition just in accord with its experimental indications. The suggestion of QD-neutrinos by the known CL mass pattern data seems artificial only until this pattern is placed against the background of an exactly degenerate pattern with mass ratios $X_n=1$. In terms of MDD-quantities, i.e. deviation of CL mass pattern from a mass-degenerate one, that suggestion looks natural[1]. (II) There is an essential connection **b**etween lepton mass hierarchies and low energy electroweak coupling constants[2]. It is remarkably supported by an observed precise connection between the fine structure constant at zero momentum transfer and the exponential $\alpha_o \equiv e^{-5}$ that is suggested by experimental evidence from CL mass-ratios. Likely answers to the mentioned above three lepton flavor physics problems a)-c) are discussed in the light of that two ideas.

In Sec.2, the Dirac-Majorana MDD-oppositeness is described by two opposite solutions of a phenomenological equation for MDD-quantities. In Sec.3, a relevant precise connection between the low energy fine structure constant and the charged lepton mass-ratio **p**arameter $\alpha_o$ is discussed and commented. In Sec.4, small QD-**n**eutrino mass scale is estimated from a drastically growing with trend to lower masses lepton mass-ratio hierarchy without inputs from the oscillation data. In Sec.5, a different estimation of the QD-neutrino mass from oscillation data is discussed. In the extended Sec.6, a discussion of results and detailed conclusions are given.

## 2. QD-neutrino mass scale from neutrino-CL MDD-opposites

---

[1] Especially in view of the puzzle of tiny neutrino mass.

[2] A relation between electron-muon mass ratio and the low energy fine structure constant is considered in the literature time and again, e.g. [3].

**1.** In accordance with the idea of neutrino-CL MDD-opposites, a particular approach to estimation of the QD-neutrino mass scale from well known experimental CL mass data is to derive an equation for the lepton (CL and neutrino) MDD-quantities $(x_n-1)$ from accurate experimental CL mass-ratio three-flavor relations by replacing the CL mass ratios $x_n$ with lepton MDD-quantities $(x_n-1)$ [2],

$$x_n^k \rightarrow (x_n^k-1), \quad n=1,2, \tag{1}$$

here k is an arbitrary real power value essentially independent of the number n=1,2. I apply this approach to the following observed nonlinear CL three flavor hierarchy relation[3] (another special CL mass-ratio relation - in Sec.5),

$$(m_\tau/m_\mu)^2 = \xi(m_\mu/m_e); \quad (x_2^k)^2 = \xi^\kappa x_1^\kappa, \quad \xi_{exp} \cong 1.37 \cong \sqrt{2}, \tag{2}$$

and obtain a generic nonlinear equation [2] for the lepton k-order MDD-quantities $(x_n^k-1)$:

$$(x_2^k-1)^2 = \xi^k(x_1^k-1), \quad k \geq 1. \tag{3}$$

The physical meaning of Eq.(3) is invariance of the lepton 'quadratic MDD-hierarchy' value $[(x_2^k-1)^2/(x_1^k-1)]$ under the exchange of CL and neutrinos: an exact quantitative analogy between the neutrino and CL patterns in terms of quadratic MDD-hierarchies.

A relation between the CL and neutrino mass ratios should be described by two opposite solutions of the equation (3) with large and small exponents, in accordance with the neutrino-CL MDD-oppositeness idea. These solutions are given by

$$X_1 \equiv m_\mu/m_e \cong \xi \exp\chi, \quad X_2 \equiv m_\tau/m_\mu \cong \xi \exp(\chi/2), \quad \chi_{exp} \cong 5, \tag{4}$$

$$x_2 \equiv m_3/m_2 = \exp(a_\kappa r), \quad x_1 \equiv m_2/m_1 = \exp(a_\kappa r^2), \tag{5}$$

---

[3] The empirical coefficient $\xi$ in (2) can be represented as an expansion in terms of the small dimensionless parameter $\alpha_{oW} \equiv 5 \exp(-5)$, see (21) below, $\xi \cong \sqrt{2}(1-\alpha_{oW}) \cong 1.367$, it agrees with the data value $\xi_{exp}$ to within 0.001.

$$a_k r \ll 1, \quad r_{exp} \cong (\Delta m^2_{sol}/\Delta m^2_{atm}),$$

for the CL and neutrino[4] mass ratios respectively (first powers). The two solutions (4) and (5) have opposite (large and small) MDD-quantities but equal quadratic MDD-hierarchies (3).

The CL solution (4) is approximately independent of the power k in Eq.(3), but the solution for the neutrino mass ratios is not, the coefficient $a_k$ in (5) is a function of k,

$$a_k = \xi^k/k. \tag{6}$$

The interesting point here is that the coefficient $a_k$ in (6) as a function of k has a unique, absolute minimum [2]:

$$a_{min} = a_{ko} = e \log\xi \cong 0.85 \tag{7}$$

at $k = k_o \equiv 1/\log\xi$, where e is the base of natural logarithms[5].

At $k=k_o$, Eq.(3) is changed to a unique form:

$$(x_2^{ko}-1)^2 = e(x_1^{ko}-1), \quad k_o=(\log\xi)^{-1} \cong 3.2. \tag{8}$$

In the present phenomenology, there are two types of physical quantities related to the neutrino mass spectrum: independent of the coefficient $a_k$ quantities, and other ones that depend on $a_k$.

It seems reasonable to assume that the physical characteristics of the actual neutrino mass spectrum should be independent of the arbitrary value k in (3), as it is in the case (2) of CL, and are related to the solution of the extreme value problem [2] for the coefficient $a_k$: the actual lepton hierarchy equation is given by (8) instead of (3), and the actual MDD-quantities of the neutrino mass spectrum are the minimal possible MDD-values given by

$$(x_2-1)_{min} \cong a_{min} r, \quad (x_1-1)_{min} \cong a_{min} r^2, \quad a_{min} \cong 0.85. \tag{9}$$

---

[4] A sequence of neutrino masses $m_1 < m_2 < m_3$ is chosen - normal ordering. An alternative solution with $x_2$ and $x_1$ interchanged in (5) - inverse ordering - is also possible with same conclusions.

[5] The value $a_{min}$ in (7) can be expressed through the parameter $\alpha_{oW}$, see footnote [3], $a_{min} \cong e(\log\sqrt{2} - \alpha_{oW}) \cong 0.85$.

The QD-neutrino solar-atmospheric hierarchy parameter

$$r \cong (x_1^2 - 1)/(x_2^2 - 1) \cong (m_2 - m_1)/(m_3 - m_2) \qquad (10)$$

is independent of $a_k$ in contrast to the relative, dimensionless-made neutrino mass-squared differences $(x_1^2 - 1) \cong 2a_k r^2$ and $(x_2^2 - 1) \cong 2a_k r$. With parameter $r$ independent of k, the minimum value of the coefficient $a_k$ in (7) means minimal possible values of both the neutrino MDD-quantities and neutrino mass ratios, as a physical condition for the actual QD-neutrino mass pattern.

The two exponential solutions of Eq.(8) with large and small exponents for the CL and QD-neutrino mass ratios are given by

$$X_1 \cong \xi \exp 5, \ X_2 \cong \xi \exp 5/2, \ \xi \cong \sqrt{2}, \qquad (11)$$

$$x_2 \cong \exp(0.85r), \ x_1 \cong \exp(0.85r^2), \ 0.85r_{QD} \ll 1. \qquad (12)$$

Since the neutrino mass-ratio coefficient $a_{ko}$ is fixed in (7) and is not small, the neutrino solution (12) leads to the definite small value of the solar-atmospheric hierarchy parameter in the present QD three-neutrino scenario:

$$r_{QD} \ll 1, \qquad (13)$$

in agreement[6] with neutrino oscillation data: $r_{exp} \ll 1$.

It should be noted that the solutions (11) and (12) for the mass ratios of the CL ($X_{1,2}$) and QD-neutrinos ($x_{2,1}$) are very different in form and value. On contrary, the corresponding solutions for the MDD-quantities of the CL [$(X_2 - 1) \cong \xi \, (\exp 5/2)$, $(X_1 - 1) \cong \xi \, (\exp 5/2)^2$] and QD-neutrinos [$(x_2 - 1) \cong 0.85r$, $(x_1 - 1) \cong 0.85r^2$] are conformable (same hierarchical patterns) and respectively large and small.

Finally, the QD neutrino mass scale, pertaining to the minimal neutrino MDD-quantities (9), is given by

$$m_\nu \cong (\Delta m^2_{atm}/1.7r)^{1/2} = (\Delta m^2_{sol}/1.7r^2)^{1/2} = \Delta m^2_{atm}/(1.7\Delta m^2_{sol})^{1/2}. \qquad (14)$$

---

[6] The same inference (13) follows also from solution (5) of the initial Eq.(3) in spite of the arbitrary power k, because of the condition (7).

With the best-fit values of the atmospheric and solar mass-squared differences from [4-6],

$$\Delta m^2_{atm} = 2.2 \times 10^{-3} \text{ eV}^2, \quad \Delta m^2_{sol} = 7.9 \times 10^{-5} \text{ eV}^2, \quad (15)$$

the neutrino mass scale is given by

$$m_v \cong 0.19 \text{ eV} \quad (16)$$

to within a few percent.

With the $3\sigma$ CL ranges [4-6] for mass-squared differences:

$$\Delta m^2_{atm} = (1.4 \div 3.3) \times 10^{-3} \text{ eV}^2, \quad \Delta m^2_{sol} = (7.1 \div 8.9) \times 10^{-5} \text{ eV}^2, \quad (17)$$

the neutrino mass scale is

$$m_v \cong (0.11 - 0.30) \text{ eV}. \quad (18)$$

These neutrino mass values are compatible with the upper limits from known data, e.g. [4-6] and [13].

**2.** All important physical characteristics of the neutrino **m**ass spectrum, (9) and (12)-(14), are expressed through one parameter $r$ which has a special, many-sided physical meaning in the present QD-neutrino scenario:

1) It describes the solar-atmospheric hierarchy parameter

$$r \cong (\Delta m^2_{sol} / \Delta m^2_{atm}). \quad (19)$$

2) The known large hierarchy of the CL mass ratios inevitably leads to small value of the neutrino oscillation solar-atmospheric hierarchy parameter $r$ by the opposite relation between neutrino and CL MDD-solutions (small versus large mass-ratio exponents) (11) and (12) in agreement with data $r_{exp} \ll 1$ in such a definite way that without this agreement, the present QD-neutrino suggestion would be falsified (see footnote [6]).

3) The parameter $r$ is a naturally small[7] phenomenological factor in the QD-neutrino mass-ratio exponents (12).

4) It is independent of the neutrino mass scale.

---

[7] As it is a measure of the deviation from mass eigenstate symmetry of QD-neutrinos, and Ref.[7].

5) It measures the large *hierarchy* of the deviations from neutrino mass degeneracy.

There is an analogy between the two empirically known lepton MDD-hierarchies[8] — between the QD-neutrino MDD-hierarchy *r* and CL MDD-hierarchy expressed by the parameter R:

$$R \equiv (X_2{}^2 - 1)/(X_1{}^2 - 1) \cong (m_\tau/m_\mu)^2/(m_\mu/m_e)^2 \cong \exp(-5) << 1,$$

$$R_{exp} \cong 0.98\, e^{-5}. \tag{20}$$

6) The condition $r \neq 0$ in (12) determines the violation of the QD-neutrino mass eigenstate symmetry by small, hierarchical neutrino mass splitting - it hints at a dynamical connection between the solar-atmospheric hierarchy parameter *r* and the QD-neutrino mass splitting interaction.

7) By the known neutrino oscillation data, the value of the hierarchy parameter *r* may be close to the quantity $\alpha_{oW}$, that is close to the value of the semiweak analogue of the low energy fine structure constant $\alpha_W \cong \alpha_{oW}$ [1],

$$r = \lambda\,\alpha_{oW}, \;\; \alpha_{oW} \equiv 5\exp(-5) \cong 0.034, \tag{21}$$

$\lambda$ is a numerical factor of order 1.

With the conditions above, the neutrino oscillation hierarchy parameter *r* should have a more basic physical meaning than the solar and atmospheric neutrino oscillation mass-squared differences have separately.

8) There is a relevant quantitative nearness between the low energy dimensionless semiweak coupling constant $\alpha_W$ and the constant $\alpha_{oW}$ :

$$\alpha_W(Q^2 \cong 0) \cong 0.031 \approx \alpha_W(Q^2 = M_Z{}^2) \cong \alpha_{oW} \cong 0.034. \tag{22}$$

The SM value $\alpha_W\big|_{Q2 \approx 0} = (\alpha/\sin^2\theta_W)\big|_{Q2 \approx 0} \cong 0.031$ is from ref.[8] in agreement with the experimental data ref.[20], the value at $Q^2 =$

---

[8] Note that these two lepton MDD-hierarchies *r* and R are known empirical quantities and differ from the 'quadratic MDD-hierarchies' which are defined in the starting Eq.(3).

$M_Z^2$ is from ref.[13]. Note that the approximate value $\alpha_{oW}$ of the dimensionless weak coupling constant turns out nearly the pole value of the running constant $\alpha_W(Q^2)$ at $Q^2 \cong M_Z^2$ in (22). With $r = \lambda \alpha_{oW}$, Eq.(22) points to a relation between the exponential factors $r$ and $\chi$ in (12) and (11),

$$r \cong \lambda \chi \exp(-\chi), \quad \lambda \approx 1. \qquad (23)$$

Relation (23) is a quantitative connection between the small and large exponents of the lepton mass ratios in the solutions (11) and (12). This relevant relation is surmised from the conditions 6), 7) and 8).

The conditions 1)−8) for the parameter $r$ in the exponents of **QD**-neutrino mass ratios are quite general. They follow from the definition of QD-neutrinos and the positive result of neutrino oscillation experiments $r_{exp} \ll 1$ if used instead of the inferred above condition $r_{QD} \ll 1$ (comp. Sec.2 of the first hep-ph in [2]).

The oppositeness relation between the neutrino and CL dimensionless MDD-quantities is substantiated by connection (23) between the small and large exponential factors $r$ and $\chi$:

$$[(m_3/m_2)^2 - 1]_{nu} \cong 1.7 \lambda \chi \xi^2 / [(m_\tau/m_\mu)^2 - 1],$$

$$[(m_2/m_1)^2 - 1]_{nu} \cong 1.7 \lambda^2 \chi^2 \xi^2 / [(m_\mu/m_e)^2 - 1]. \qquad (24)$$

These connected neutrino and CL MDD-quantities obey the hierarchy equation (8).

The large CL mass ratios $(m_\tau^2/m_\mu^2)$ and $(m_\mu^2/m_e^2)$ have direct physical approximate meaning of MDD-quantities for the CL, while **t**he inverse small CL mass ratios $(m_\mu^2/m_\tau^2)$ and $(m_e^2/m_\mu^2)$ have only **i**ndirect MDD-meaning: they determine the small MDD-quantities of the QD-neutrinos by relations (24).

From (24) and (20), it follows

$$r_{QD} \cong (x_1^2 - 1)/(x_2^2 - 1) \cong 5\lambda (m_\tau/m_\mu)^2 / (m_\mu/m_e)^2 \equiv 5\lambda \ R. \qquad (25)$$

So, the two observable lepton MDD-hierarchies $r$ and R are related to one another: the ratio of these hierarchies in (25) is $r/R \cong 5\lambda$. With that, the well known very large[9] CL mass-ratio squared hierarchy $R \cong 1/151$ from **(20)** leads once more to an estimation of a large hierarchy of the solar and atmospheric neutrino oscillation mass-squared differences, $r_{QD} \cong \lambda/30$.

Integer 5 as a likely discrete symmetry characteristic of the QD-neutrino mass matrix with approximate bilarge mixing, induced radiatively in the context of broken supersymmetry, is considered in [9]. The possibility of baryogenesis with QD-neutrinos, $m_v \geq 0.1$ eV, is considered in [10]. QD-neutrino masses are discussed within different seesaw approaches in e.g. [11], [17] and [19].

### 3. The CL parameter exp(-5) and low energy fine structure constant

**1.** Why the exponential $e^{\pm 5}$? There are at least two possible answers to this unavoidable question: 1) The experimental evidence of the exponential exp($\pm 5$) in the CL mass ratios (11), in the low energy semiweak coupling constants (21),(22), in the MDD-hierarchies (20) and (21) and neutrino mass ratios (12) is a set of coherent coincidences. 2) The appearance of the exponential $e^{\pm 5}$ is related to new flavor physics where it plays a universal dimensionless physical parameter. In the interesting second case, further relations between the low energy gauge coupling constants of the electroweak theory [12] and the exponential $e^{-5}$ should be expected. It is observed below that there is a definite explicit relation between the precision data

---

[9] Unlike the MDD-quantities themselves, both the large and small values of the *ratios* of MDD-quantities describe 'large MDD-hierarchies'.

value of the fine structure constant and just this CL mass-ratio parameter $e^{-5}$.

1) The approximate relation

$$\alpha \cong \alpha_o \equiv e^{-5}, \qquad (26)$$

where $\alpha \cong 1/137$ is the empirical value [13] of the fine structure constant, is correct to within ~8%. Approximation (26) is prompted by the lepton mass-ratio indications and the relevant relation (22) for the semiweak constant $\alpha_W \approx 5\exp(-5)$ plus the approximate empirical relation [13] for the Weinberg mixing angle $\theta_W$ of the electroweak theory: $\sin^2\theta_W \cong 0.2$.

Relation (26) is a remarkable one. With (11)-(23), and the electroweak theory [12] relation $\alpha_W = \alpha/\sin^2\theta_W$, approximate relations between lepton mass ratios and the empirical CL mass-hierarchy parameter $\alpha_o$ are given by

$$\alpha \approx \alpha_o, \;\; \sin^2\theta_W \approx 1/\log\alpha_o^{-1} = 0.2, \;\; \alpha_W \approx \alpha_{oW} \equiv \alpha_o \log\alpha_o^{-1}, \qquad (27)$$

$$m_\mu/m_e \cong \sqrt{2}/\alpha_o, \;\; m_\tau/m_\mu \cong \sqrt{(2/\alpha_o)}, \;\; R \cong \alpha_o, \qquad (28)$$

$$(m_3/m_2-1)_{nu} \cong 0.85r, \;\; (m_2/m_1-1)_{nu} \cong 0.85\,r^2, \;\; r = \lambda\,\alpha_{oW}, \qquad (29)$$

$$\alpha_{oW} \equiv 5e^{-5}, \;\; \lambda \approx 1,$$

where $r$ and R are the MDD-ratios (hierarchies) of respectively neutrino and CL masses (10), (20) and (21),(23).

Relations (27)-(29) are correct to within ~(1-8)%. The empirical parameter $\alpha_o$ approximately unites low energy electroweak gauge coupling constants with the corresponding lepton mass-ratios[10]; five independent dimensionless coupling constants of the low energy electroweak lepton interactions, with the electromagnetic field and $W^\pm$-field (gauge interactions with coupling constants $\alpha$ and $\alpha_W$) and scalar Higgs-field interactions ($f_e$, $f_\mu$ and $f_\tau$ — coupling constants with the scalar field of the electron, muon and tauon respectively)[11],

$$f_\mu \cong (\sqrt{2/\alpha_o}) f_e, \quad f_\tau \cong \sqrt{(2/\alpha_o)} f_\mu, \tag{30}$$

are united via the parameter $\alpha_o$; one coupling constant with the scalar field is left free (e.g. $f_e$, i.e. electron mass $m_e$).

---

[10] With the estimation of neutrino mass scale $m_\nu$ in (18), the ratio of that scale to the electron mass $m_e$ should be $(m_\nu/m_e) = \eta\,\alpha_o^3$, where the coefficient $\eta$ is $\eta \cong (0.7 \div 1.9)$ at the $3\sigma$ C.L.

2) Observe now, as a *fact* with $(\alpha^{-1})_{Data} \cong 137.036$ from [13], that the following extension of the approximate relation (26),

$$\exp 2\alpha \, \log(\exp\alpha /\alpha) \cong \log(1/\alpha_o), \tag{31}$$

is correct to within $\sim 3 \times 10^{-6}$. This large increase in accuracy is achieved not by consecutive adjusting terms, but by the addition of two close to unity and clearly related exponential factors $(\exp\alpha)$ and $(\exp\alpha)^2$ to the relation (26) without fine tuning[12].

Considering (31) as an equation for the unknown $\alpha$, the solution is

$$\alpha \cong 1/137.0383, \quad (\alpha -\alpha_{Data})/\alpha_{Data} \cong -1.7 \times 10^{-5}, \tag{32}$$

compared to $[\alpha_o-\alpha_{Data}]/\alpha_{Data} \cong -0.08$.

Another very interesting definite indication from Eq.(31) is that, after exponentiation,

$$(\exp\alpha /\alpha)^{\exp 2\alpha} \cong \exp 5, \tag{31'}$$

the difference between the right and left sides in (31') is equal to $(\alpha/\pi)$ to within $\sim 2 \times 10^{-3}$ — an impressive relative accuracy.

---

[11] For the CL: $m_\ell = f_\ell \langle\varphi\rangle$, $\ell$ = e,$\mu$,$\tau$, $\langle\varphi\rangle$ is the vacuum expectation value of the scalar field $\varphi$.

[12] Equation (31) is a phenomenological one, presumably derivable in a finite nonperturbative theory of lepton masses and interactions if starting with the primary value $\alpha_o = \exp(-5)$ of the fine structure constant.

3) So, a further extended relation

$$\exp 2\alpha \, \log(\exp\alpha /\alpha) \cong \log(1/\alpha_o - \alpha/\pi) \qquad (33)$$

between the source-value $\alpha_o$ and the highly accurate data value of the fine structure constant [13],

$$(1/\alpha)_{Data} = 137.03599911(46), \qquad (34)$$

is satisfied to within $\sim 6 \times 10^{-9}$.

If considering (33) as an equation for the unknown $\alpha$, the solution is

$$\alpha \cong 1/137.0359948, \quad (\alpha - \alpha_{Data})/\alpha_{Data} \cong 3.1 \times 10^{-8}. \qquad (35)$$

It differs from the central data value of the fine structure constant (34) by about 10 S.D.

Note that the additional term $\alpha/\pi$ on the right side of (33) is like a "perturbation term" added to the "nonperturbative" basic relation (31).

The sequence of consecutive empirical relations of growing accuracy (26), (31) and (33) between the data value of the fine structure constant $\alpha_{Data}$ and the source value $\alpha_o$ supports the independent indications in Sec.2 that the exponential $\alpha_o$ should be a universal physical parameter in the phenomenology of lepton flavor physics.

4) Finally, a possible second "perturbative" term (e.g. $\alpha^2/4\pi$) raises the accuracy of relation (33) by about one order of magnitude, but transforms it into a highly accurate nonlinear equation for the unknown $\alpha$:

$$\exp 2\alpha \, \log(\exp\alpha /\alpha) = \log(1/\alpha_o - \alpha/\pi + \alpha^2/4\pi), \qquad (36)$$

or, comp. (31'),

$$(\exp\alpha /\alpha)^{\exp 2\alpha} + \alpha/\pi - \alpha^2/4\pi = \exp 5. \qquad (36')$$

Indeed, the solution of Eq.(36) is given by

$$\alpha \cong 1/137.03599901, \quad (\alpha - \alpha_{Data})/\alpha_{Data} \cong 0.7 \times 10^{-9}. \qquad (37)$$

This solution $\alpha$ of the phenomenological equation (36) precisely agrees with the central data value of the fine structure constant at zero momentum transfer $\alpha_{Data}$ in (34), to within ~0.2 S.D.

Equation (36) is a virtual accurate equation for the fine structure constant $\alpha$ at zero momentum transfer with the exponential exp5 on the right side of Eq.(36') as the source of the precise numerical solution (37). A more compact form of Eq.(36) is given by

$$(\exp\alpha/\alpha)^{\exp 2\alpha} + (\alpha/\pi)\exp(-\alpha/4) = \exp 5. \qquad (36'')$$

The solution of this equation is $\alpha \cong 1/137.03599900$.

**2.** The interesting question of where does the specific **n**umerical value of the fine structure constant at zero momentum transfer come from has a likely answer. With the relations above, a unique connection between the fine structure constant $\alpha$ at $Q^2=0$ and the dimensionless parameter $\alpha_o$ is possible, at least to within a few S.D. If the true numerical value of the fine structure constant at zero momentum transfer is reduced to integer 5 (alpha-genesis[13]) through the source value $\alpha_o \equiv e^{-5}$, all gauge coupling constants of the Standard Model might be expressed through the parameter $\alpha_o$ by renormalization group equations and Grand Unification [14].

---

[13] As a finite number, the value $\alpha_o$ uniquely determines the empirical value $\alpha_{Data}$ and with the QED renormalization group equations - all the further growing values of the fine structure constant at higher momentum transfers.

The minimal observable value of the fine structure constant $\alpha$ at the particular momentum transfer $Q^2=0$ is a very special physical quantity because of its role in nonrelativistic quantum mechanics of bound states enabling Life and Consciousness. Its unique relation to the exponential $\alpha_o \equiv e^{-5}$ is a possible solution to the anthropic principle problem for the fine structure constant in our universe.

**3.** We should point out that there is also another reason why the relations (27)-(29) are particularly interesting: if they approximately represent essential primary connections in particle physics, they afford a special answer to the important question of how is it that two extra lepton flavors, muon and **t**auon, are needed in the low energy physics region $E < m_e/\alpha^2$. With $\alpha_o$ as a universal parameter in flavor physics, the two extra CL flavor generations beyond the first one would be unobservable at $\alpha_o \to 0$; with finite electron mass $m_e$, the $m_\mu$ and $m_\tau$ would be extremely large and the neutrino mass would be $m_\nu = 0$ and exactly degenerate, but the electroweak interactions of the electron would disappear. So, in flavor phenomenology with the universal parameter $\alpha_o$, the muon and tauon with known finite masses are necessary for the electron-flavor generation particles to have bound states.

**4.** By Eqs.(11), (20) and (36), a low energy relation appears between flavor-physics and QED-physics via the fine structure constant $\alpha$, CL mass and mass-ratio hierarchies and the universal **p**arameter $\alpha_o$. This parameter $\alpha_o$ has a dual functional meaning: from (36) it is the source of the fine structure constant value in QED, and at the same time from (20) it is the source of the CL mass-ratio hierarchy R in flavor physics, $R \cong 0.98\,\alpha_o$. The gauge symmetry of QED determines conservation of electric charge and the interaction of the electron with electromagnetic field, but not the value of the elementary charge $\varepsilon = (4\pi\alpha)^{1/2}$. Its absolute value is related here to the parameter $\alpha_o$ and further to the CL MDD-quantities $(X_1^2-1)$ and $(X_2^2-1)$ and their hierarchy R. All these four dimensionless low energy physical quantities may have a common origin - the parameter $\alpha_o$. The chain $R \longleftrightarrow \alpha_o \longleftrightarrow \alpha$ seems a relevant hint from experiments on a solution to the old problem of what is the physical meaning of the electron flavor copies; what kind of physical information their mass pattern carries and how to unite the three copies into one flavor system characterized by its tri-large pattern of mass and mass-ratio hierarchies, (11),(20).

**5.** The fine structure constant value and the absolute value of the elementary charge at $Q^2=0$ are encoded in the ripples (hierarchies) on a pattern of mass-degenerate CL copies[14] -- a fundamental physics aspect of the low energy phenomenology.

## 4. QD-Majorana-neutrino mass from drastic lepton mass-ratio hierarchy

In the QD-neutrino scenario, the lepton mass spectrum contains 4 mass-degenerate Majorana mass levels: three two-fold exactly-mass-degenerate[15] CL mass levels $m_\tau$, $m_\mu$ and $m_e$ (three carrying charge Dirac states) plus one three-fold Quasi-Degenerate Majorana-neutrino mass level $m_\nu$. The three CL mass levels are highly hierarchical, with the hierarchy-rule approximately described by Eq.(28) in terms of powers of the universal parameter $\alpha_o = e^{-5}$. Can this high CL mass-ratio hierarchy be extended so to include the fourth comparatively very low QD-neutrino mass level $m_\nu$? An affirmative answer is indeed possible

---

[14] In accordance with the solution (11) for the CL mass ratios, these ratios can be described by one formula
$$X_n \cong \sqrt{2} \, \exp(5/2^n) \equiv \sqrt{2} \, (\alpha_o^{-1})^{(1/2^n)}, \quad X_n = m_{n+1}/m_n,$$
$n = 0,1,2$ for $e$, $\mu$ and $\tau$ respectively. Obviously, it can be rewritten in the form
$$X_n \cong \sqrt{2} \, \exp(\chi_i/2^{n+i}), \quad \chi_i = 5 \times 2^i,$$
with the index i being independent of the index n. Just, the electron number is an arbitrary integer 'i', but the flavor mass hierarchies are independent of the electron number.

[15] This exact mass-degeneracy (symmetry) cannot be broken down so far as there are no interactions that violate electric charge conservation, in contrast to conservation of lepton charge.

in the QD-neutrino scenario: factorial hierarchy, which extends the sequence in (28):

$$m^2_{\ell+1}/m^2_\ell \cong \alpha_o{}^{\ell!}/2. \qquad (38)$$

The notations are $m_1 = m_\tau$, $m_2 = m_\mu$, $m_3 = m_e$ and $m_4 = m_\nu$. The inclusion of the mass level $m_\nu$ into the CL mass-ratio hierarchy is enabled by a complementary remote view at the QD-neutrino mass pattern.

The factorial mass-ratio hierarchy between four lepton mass levels (38) is represented by three terms

$$m^2_\mu/m^2_\tau \cong \alpha_o/2, \ m^2_e/m^2_\mu \cong \alpha_o{}^2/2, \ m^2_\nu/m^2_e \cong \alpha_o{}^6/2. \qquad (39)$$

Hence, the absolute value of the QD-neutrino mass scale is given by

$$m_\nu \cong \alpha_o{}^3 \, m_e/\sqrt{2} \cong 0.11 \text{ eV}. \qquad (40)$$

Because of the high power of the constant $\alpha_o$ in (40), a change of the value $\alpha_o$ to the exact value of the low energy fine structure constant $\alpha$ may lead to an increased neutrino mass scale

$$m_\nu \cong \alpha^3 \, m_e/\sqrt{2} \cong 0.14 \text{ eV}. \qquad (41)$$

The estimations (40) and (41) are compatible with the cosmological neutrino bounds [16] for QD-neutrinos, $m_\nu < 0.14$ eV at 95% C.L.

In summery, a drastically growing in the direction of smaller masses mass-ratio hierarchy of the four lepton mass levels $m_\tau$, $m_\mu$, $m_e$ and $m_\nu$ is another way to describe the extreme relative smallness of the QD-neutrino mass scale $m_\nu$ within the considered lepton MDD-phenomenology without use of neutrino oscillation data, in contrast to Sec.2.

## 5. QD-neutrino mass scale from Koide CL mass formula

1. Another example of a particularly interesting low energy relation between CL mass ratios is the Koide formula [21]:

$$(m_e + m_\mu + m_\tau) = 2/3 \, (\sqrt{m_e} + \sqrt{m_\mu} + \sqrt{m_\tau})^2 \qquad (42)$$

It is considered as mysterious [21]. Indeed, it looks a mysterious coincidence ($\sim 10^{-5}$ - for the central value of the tauon data mass $m_\tau \cong 1777$ MeV) between the physical lepton mass ratios, not invariant under the renormalization group transformations.

With the notations above for the CL mass ratios $X_1 = m_\mu/m_e$, $X_2 = m_\tau/m_\mu$, rewrite (42) in the form

$$(1 + X_1 + X_1 X_2)\,3/2 = [1 + \sqrt{X_1} + \sqrt{(X_1 X_2)}]^2. \qquad (43)$$

In Sec.2, 'bare' values of the CL mass ratios are introduced,

$$X_{o1} = \sqrt{2} \, e^5, \quad X_{o2} = \sqrt{2} \, e^{5/2}. \qquad (44)$$

With these values, the relation

$$(1 + X_{o1} + X_{o1} X_{o2})\,3/2 \cong [1 + \sqrt{X_{o1}} + \sqrt{(X_{o1} X_{o2})}]^2 \qquad (45)$$

is correct to within $\sim 4 \times 10^{-3}$. It seems that the approximate relation (45) between the bare values of CL mass ratios[16] may be the right basis of the formula (42). Here, there are no problems with renormalization group invariance since the bare mass ratios are numbers. Because of the fact that the relative deviations of the physical CL mass ratios from their bare values (44) are of the order $\sim 10^{-2}$, the necessary 'finite radiative corrections' to the bare values of the CL mass ratios should have the right size to cancel the remaining small inequality between the left and right sides of the relation (45), and thus render it more accurate. In that light, the high accuracy of the Koide formula, especially <u>if</u> its prediction of the tau-lepton mass $m_\tau = 1776.97$ MeV will be confirmed, looks more probable (less mysterious).

---

[16] Accurate relations between physical mass ratios seem more probable if they are supported by corresponding relations between bare mass ratios, comp. (2),(3).

2. In any case, the Koide formula is an accurate empirical CL three-flavor mass-ratio relation. In accordance with neutrino-CL MDD-oppositeness idea, the mass scale $m_\nu$ of QD-neutrinos may be inferred from the relation (43) in analogy with the case of CL mass-ratio relation (2), in this case by the substitution:

$$X_{2,1}^2 \rightarrow (x_{1,2}^2 - 1) \cong (\Delta m_{sol}^2; \ \Delta m_{atm}^2)/m_\nu^2 \qquad (46)$$

for n=1,2 respectively, comp.(1). An equation for QD-neutrino MDD-quantities is deduced from (43) in the form:

$$[1+(x_2-1)+(x_1-1)(x_2-1)]3/2 = [1+\sqrt{(x_2-1)}+\sqrt{(x_1-1)(x_2-1)}]^2. \qquad (47)$$

With the condition $m_\nu \geq 0.1$ eV, and the neutrino mass-squared differences from oscillation data (17), an approximate relation for the QD-neutrino mass scale follows from (47)

$$m_\nu^2 \cong 8 \, \Delta m_{atm}^2. \qquad (48)$$

With neutrino oscillation data (17), the QD-neutrino mass scale values at $3\sigma$ are

$$m_\nu \cong (0.11 - 0.16) \text{ eV}, \qquad (49)$$

and the best fit value is

$$(m_\nu)_{bf} \cong 0.12 \text{ eV}. \qquad (50)$$

The estimation of the QD-neutrino mass scale in Sec.2, Eq.(18), is determined by both atmospheric and solar oscillation mass-squared differences ($\Delta m_{atm}^2$, $\Delta m_{sol}^2$); the estimation of this quantity in Sec.4, Eqs.(40),(41), is determined by lepton mass-ratio hierarchy (38) and is independent of the neutrino oscillation data; finally, the estimations of $m_\nu$ in (48)-(50) are determined almost exclusively by the atmospheric neutrino oscillation mass-squared difference $\Delta m_{atm}^2$. Remarkably, all three independent estimations of the QD-neutrino mass scale from the experimental data are in fair agreement with each other.

## 6. Discussion and conclusions

Unlike the one-generation Standard Model, there is no well established theory of flavor particle physics as yet. In this paper, a definite low energy lepton mass-flavor phenomenology is considered, it is based on explicit ideas and a system of experimental evidence, and it is testable by new coming neutrino mass and oscillation data. Though not directly related to mainstream particle frontier theory with symmetry probes at high energy scale, that phenomenology relates some known low energy lepton flavor-physics problems, defines some new ones and enables quantitative verifiable estimations of important neutrino mass and oscillation quantities. It is a valid generic approach at a frontier physics sector.

An extreme value problem for the neutrino $(x_n-1)$-MDD-quantities leads from the data indicated Eq.(3), that includes an arbitrary parameter, to the definite nonlinear lepton mass-ratio equation (8). It has two solutions with large and small exponents, that may describe the CL and neutrino MDD-patterns and, with data inputs, predicts small QD-neutrino masses, small solar-atmospheric hierarchy parameter $r \ll 1$ and a connection between the two observable large lepton MDD-hierarchies $r$ and R, Sec.2. The MDD-hierarchy quantities R and $r$ of the CL and QD-neutrinos are unique analogous three-flavor characteristics of three generation lepton physics in the present paper.

Precise empirical relations between the fine structure constant at momentum transfer $Q^2=0$ and the integer 5 are observed. The empirical functional dependence on integer 5 of the MDD $(x_n-1)$ quantities of the CL and QD-neutrinos approximately resembles[17] that of the low energy electroweak coupling constants in the order $(1/\alpha)$ and $\alpha_W$ respectively.

In the present low energy lepton flavor phenomenology, the particle MDD patterns, as patterns of violation from exact flavor symmetries, are thought more primary concepts than the symmetries themselves: hierarchies carry the physical information about the low energy electroweak particle charges; they may suggest physical meaning to particle flavor copies, while exact mass-degeneration symmetry is only a background[18].

### --Conclusions--

The following conclusions are based on a system of approximate experimental evidence from lepton mass-flavor and electroweak physics. The conclusions are detailed since it should be made clearer that they are supported by the whole evidential system, rather than by particular data pieces.

*1) The idea of neutrino-CL MDD-oppositeness* - one of the two main guiding ideas in this work. It is represented in Sec.2 by two exponential lepton mass-ratio solutions for lepton particles with respectively large and small exponents (11) and (12). A relation between two highly hierarchical pairs of MDD-quantities of the CL $[(X_1{}^2-1)_{CL} \cong (m_\mu/m_e)^2, \ (X_2{}^2-1)_{CL} \cong (m_\tau/m_\mu)^2]$ and QD-neutrinos $[(x_2{}^2-1)_{nu} \cong 1.7\,r, \ (x_1{}^2-1)_{nu} \cong 1.7\,r^2]$ follows from these solutions with the consistency condition $r \ll 1$.

---

[17] "…in the description of nature, one has to tolerate approximations, and that even work with approximations can be interesting and can sometimes be beautiful" - P. A. M. Dirac, Scientific autobiography, in *History of 20th Century Physics*, NY (1977).

[18] The physical content of lepton flavor hierarchies may imply a fundamental aspect of low energy phenomenology, also Sec.3.

*2) A source-value of the fine structure constant $\alpha_o \equiv e^{-5}$ is* unearthed in Sec.3. The parameter $\alpha_o$ determines the precise data value of the fine structure constant at zero momentum transfer by the virtual equation (36). It is certainly pertinent to the discussed lepton mass-ratio phenomenology. This conclusion is supported by a consistent system of experimental evidence, see e.g. (27)-(29), (39) and below.

*3) With the new basic parameter $\alpha_o$ in flavor physics, the muon and tauon flavor counterparts of the electron are unavoidable.* If the approach $\alpha_o \to 0$ were made, the lepton electroweak interactions would **d**isappear[19]. With this universal parameter[20] $\alpha_o$ in low energy flavor physics, a physically meaningful limiting case with only one flavor generation cannot be imagined; without the muon, tauon and probably QD-neutrinos there would be no bound states of the electron. From neutrino-CL MDD-oppositeness, the neutrinos cannot be massless just because the muon and tauon cannot be infinitely heavy. The finite electron mass $m_e$ and finite small value $\alpha_o$ provide a quantitative answer to the actual question of why the muon and tauon masses are large $\gg m_e$ but not very large $(m_\mu, m_\tau) < m_e/\alpha^2$, while the neutrino masses are small $m_\nu \approx \alpha^3 m_e$, but not zero, e.g. (39).

---

[19] At the zero approximation in the fine structure constant $\alpha = \alpha_o = 0$, with $m_e$ finite, we get $m_\tau, m_\mu = \infty$ and $m_\nu = 0$, and no electroweak interactions. A very small change of the fine structure constant from $\alpha = 0$ to $\alpha = \alpha_{Data}$ would generate a infinitely large decrease of the muon and tauon masses to their data values together with a very small increase of the neutrino mass scale from zero to $(m_\nu)_{QD} > 0$ (for illustration only).

[20] All considered in this paper dimensionless quantities — CL and QD-neutrino mass ratios, MDD-values and MDD-hierarchies, the low energy electroweak coupling constants $\alpha$ and $\alpha_W$ and the three CL Yukawa coupling constants — are nearly expressed through the parameter $\alpha_o$, comp.(27) -(29), (36), (39), (51) - (57). It is simpler to read an indication of new physics than to accept so many coincidences.

*4) Both the CL and QD-neutrino MDD-ratios (10) and (20) describe large[21] MDD-hierarchies, they are experimentally known quantities*. The suggestion that the solar-atmospheric hierarchy parameter $r$ is related to the low-energy dimensionless coupling constant $\alpha_W$ should be commented. Consider three very different estimations of the hierarchy parameter $r$:

i) $\quad r = \lambda \alpha_{0W} \cong \lambda \alpha_W(Q^2=M_Z^2) \cong 0.034 \lambda, \ \lambda \cong 1.$ (51)

ii) With the SM prediction [8], $\sin^2\theta_W|_{Q2=0} \cong 0.2383$, it follows

$\quad r = \lambda_1 \alpha_W(Q^2=0) = \lambda_1 (\alpha/\sin^2\theta_W)|_{Q2=0} \cong 0.031 \lambda_1, \ \lambda_1 \cong 1.$ (52)

iii) By shifting the value $\alpha_0$ to $\alpha = \alpha_{Data}$ in (27), the parameter $r$ is given by

$\quad r \cong \lambda_2 [\alpha \ln(1/\alpha)]_{Q2=0} \cong 0.036 \lambda_2, \ \lambda_2 \cong 1.$ (53)

These estimations are examples of different concrete interpretations of the suggested connection between the solar-atmospheric hierarchy parameter $r$ and the dimensionless weak coupling constant $\alpha_W$. The estimations (51),(53) are in better agreement with the best fit value of ref.[4]. In any case, if the suggested relation has phenomenological meaning, it should be $r/\alpha_{0W} \approx 1$.

*5) Essential connections between the lepton mass hierarchies and the low energy electroweak coupling constants* - the second one of the two guiding ideas in this paper. It is a postulate which ultimately leads to the emergence of the universal parameter $\alpha_o$ in the analysis of the dimensionless experimental flavor and electroweak lepton quantities in this paper. It attaches common sense to several factual 'Why' questions concerned with $\alpha$—related values of the experimental lepton flavor and neutrino oscillation patterns: why extra CL beyond the electron are attainable in the low energy region $E < m_e/\alpha^2$ of particle physics; why the experimental CL mass ratios and their hierarchy appear expressed through one factor $\alpha \cong \alpha_o = e^{-5}$ and why the fine structure constant $\alpha$ at $Q^2 = 0$ is determined just by this experimental factor of the CL mass ratios; why the small value of the solar-atmospheric hierarchy parameter $r$ from the neutrino oscillation data may be close to the low energy weak coupling constant $\alpha_W$; why the large experimental MDD-hierarchy $r$ from the neutrino oscillation data resembles the known CL mass-ratio hierarchy; why the QD-neutrino mass scale is so much smaller than the electron mass $m_\nu < \alpha^3 m_e$. That idea is described in Sec.3 by the precise Eq.(36), and in Sec.2 by two approximate pairs (with conformable hierarchy structures) of relations between the large and small lepton MDD-quantities and the low energy electroweak coupling constants $\alpha \cong \alpha_o$ and $\alpha_W \cong \alpha_{oW}$:

$$(X_2^2 - 1)_{CL} \cong 2\,(\alpha_o^{-1}), \quad (X_1^2 - 1)_{CL} \cong 2\,(\alpha_o^{-1})^2,$$

$$(x_2^2 - 1)_{nu} \cong 1.7\,(\alpha_{oW}), \quad (x_1^2 - 1)_{nu} \cong 1.7\,(\alpha_{oW})^2, \quad \alpha_{oW} \equiv 5\alpha_o. \qquad (54)$$

---

[21] Unlike the MDD-quantities themselves, large and small values of the *ratios* of MDD-quantities, by an obvious reason, describe the same 'large MDD-hierarchies' (a generic characteristic of lepton mass patterns) in contrast to 'order 1 MDD-hierarchies' in case of near geometrical mass patterns (a probable characteristic of flavor quark mass patterns [18]).

It looks like the known CL MDD-pattern in (54) suggests the hypothetical QD-neutrino pattern[22] (lower pattern in (54)).

---

[22] The suggestion of QD-neutrinos by the known CL mass pattern seems artificial only until this pattern is placed against the background of exactly degenerate pattern with mass ratios $X_n = 1$. In terms of MDD-quantities, i.e. deviation of CL mass pattern from a mass-degenerate one or deviations of mass ratios from 1, that suggestion looks natural, especially on account of the prolonged quest for neutrino mass pattern. It is indeed natural to count very large and very small deviations $(x_n - 1)$ as possible ones within one united system of massive lepton particles.

As an inference from the relations (54), the most concise description of the lepton flavor-electroweak connection is given below by two approximate relations (56),(57) between the unique lepton three-flavor MDD-hierarchies $r$ and R, on the one hand, and the EW coupling constants $\alpha_W$ and $\alpha$, on the other hand.

**O**bserve also four low energy approximate relations

$$m_\tau \cong \alpha <\varphi>, \; m_\mu \cong \alpha \, (\alpha_o/2)^{1/2} <\varphi>, \; m_e \cong (\alpha/2)\alpha_o^{3/2} <\varphi>,$$

$$m_W \cong (\pi\alpha_{oW})^{1/2} <\varphi>, \qquad\qquad (55)$$

accurate to within $\sim(1\div3)\%$ (comp. (55) with (28)). Here $<\varphi> \cong 246$ GeV is the vacuum expectation value of the Higgs scalar field[23]. The empirical relations in (55) determine the Yukava coupling constants of the physical CL masses, and W-boson mass $m_W$. So, a meaningful system of experimental evidence in favor of the postulate '5)' is produced with parameter $\alpha_o$ as a probable link.

The lepton masses and EW charges are independent empirical parameters in the one-generation EW theory [12]. In the present paper, flavor copies enable a connection between the mass spectra of CL and neutrinos with EW theory charges: these charges are encoded in the lepton two-flavor MDD-quantities and three-flavor MDD-hierarchies. The hierarchies of lepton mass copies may attach the missing physical meaning to particle flavor in the low energy EW physics.

*6) Lepton MDD-hierarchies and three flavors.* In the framework of the electroweak theory, the empirical fact of three sequential lepton flavors is a challenging mystery - there is no

---

[23] As an example of other sort of empirical relations, the heaviest and lightest masses of the low energy lepton mass spectrum are given by

$$m_\tau \cong \alpha <\varphi>, \quad m_\nu \cong \alpha \, E_r \cong 0.10 \, eV,$$

where $E_r$ is the Ridberg energy [13] from atomic physics $E_r = m_e \alpha^2/2$.

known essential dependence of electroweak properties on masses of the different generation particles. One flavor is very natural in the EW theory, and so why are three flavors needed? In the present phenomenology, the otherwise independent value of the elementary charge $\varepsilon = (4\pi\alpha)^{1/2}$ is encoded in the three CL flavor-mass-pattern hierarchy R. Proper connections between the two three-flavor MDD-hierarchy quantities of the neutrinos $r$ and CL R and two low energy electroweak coupling constants ($\alpha \cong \alpha_o$) and ($\alpha_W \cong \alpha_{oW}$) [24] are observed:

$$r = \lambda \alpha_{oW}, \ R = \lambda' \alpha_o; \ \lambda, \ \lambda' \cong 1. \tag{56}$$

The first relation is a discussed in the text suggestion, while the second one is a fact $\lambda'_{exp} \cong 0.98$.

An interesting parallelism of the two CL and QD-neutrino chains,

$$R \longleftrightarrow \alpha_o \longleftrightarrow \alpha,$$

$$r \longleftrightarrow \alpha_{oW}(\alpha_o) \longleftrightarrow \alpha_W, \tag{57}$$

where the right and left ends are physical observables and the middle links are definite numbers, suggests relations between essential lepton flavor and electroweak quantities. This parallelism continues the emphasized in Sec.2 analogy between important three-flavor neutrino and CL MDD-hierarchy parameters $r$ and R.

The analogous relations in (57) are unique only in the case of three particle generations. There would be no MDD-hierarchies in case of two flavors, whereas in case of more than three flavors there would be more than two MDD-**h**ierarchies (not unique), while there are only two independent EW gauge coupling constants $\alpha$ and $\alpha_W$.

---

[24] The constant $\alpha_{oW}$ is analogous to the source value $\alpha_o$ of the fine structure constant $\alpha$. The difference is that $\alpha_o < \alpha(Q^2=0)$, while $\alpha_{oW} > \alpha_W(Q^2=0)$. Since the running coupling constant $\alpha_W(Q^2)$ increases in the space between $Q^2=0$ and $Q^2 \approx M_Z^2$, the condition $\alpha_W(Q_1^2) = \alpha_{oW}$ may be realized at some value $Q_1^2 > 0$ which in fact, according to (22), is indeed realized at $Q_1^2 \approx M_Z^2$.

The approximate values $\alpha_o$ and $\alpha_{oW}$ of the electroweak coupling constants appear as links between electroweak and flavor physics: they are sources of the lepton MDD-hierarchies in flavor-physics, and they are sources of the numerical values of coupling constants in EW-physics.

*7) Drastically growing in the direction of smaller masses mass-ratio-hierarchy of four lepton mass levels* $m_\tau$ *,* $m_\mu$ *,* $m_e$ *and* $m_\nu$ is considered in Sec.4. It employs a complementary aspect of the QD-neutrino mass pattern as a degenerate one, from comparison with CL large mass ratios. It is another way to describe the extreme smallness of the QD-neutrino mass scale $m_\nu$ from neutrino-CL MDD-oppositeness suggestion and CL mass-ratio data.

*8) The main verifiable results are:*

(a) Neutrino mass pattern is a special quasi-degenerate one[25]; this result is partially supported by neutrino oscillation data $r_{exp} \ll 1$ which are in agreement with the nontrivial inference of small solar-atmospheric hierarchy parameter $r \ll 1$ as a consistency condition without use of oscillation data, comp. (5), (7) and (12).

(b) Estimations of QD-neutrino mass scale are $m_\nu \cong (0.11-0.30)$eV and $m_\nu \cong (0.11-0.16)$eV from oscillation data, and $m_\nu \cong (0.11-0.14)$eV from drastic lepton mass-ratio hierarchy; the fairly good agreement between these three independent estimations of the QD-neutrino mass scale is an interesting quantitative result from neutrino oscillation and CL mass-hierarchy data in the light of the neutrino-CL MDD-oppositeness idea.

---

[25] A test of QD neutrino mass type is in the sensitivity region of the present and discussed experiments, see [15] and [16].

(c) The parameter $\alpha_{oW} \equiv 5\alpha_o$ is a probable link between QD-neutrino flavor physics (neutrino mass-ratios, solar-atmospheric-MDD-hierarchy parameter $r$) and weak interaction physics (a connection between $\alpha_{oW}$ and $\alpha_W$).

(d) The parameter $\alpha_o \equiv e^{-5}$ links CL flavor physics (CL mass ratios $X_1$ and $X_2$, MDD-hierarchy parameter R) with EW physics (precise connection between $\alpha_o$ and $\alpha$, CL Yukawa coupling constants); the chain $R \longleftrightarrow \alpha_o \longleftrightarrow \alpha(Q^2=0)$ is an observed hint from experiments on a probable answer to the question of what sort of new physics may render the known three sequential CL copies (e, $\mu$, $\tau$) into one CL flavor system specified by the data pattern of tri-large mass and mass-ratio hierarchies: $(X_1, X_2, R^{-1})_{exp} \gg 1$.

The results (a)-(c) — QD-neutrinos, neutrino mass scale $m_\nu$, small deviation from 1 of the coefficient $\lambda$ in the relations $r/R \cong 5\lambda$ and $\lambda = r/\alpha_{oW}$ — are testable by coming new accurate neutrino mass and oscillation data. The special CL flavor-EW result (d) is in apparent agreement with known experimental data.